\documentclass{emulateapj}
\usepackage{apjfonts}
\usepackage{graphicx}

\def\kms{\ifmmode{\rm km\thinspace s^{-1}}\else km\thinspace s$^{-1}$\fi}

\shortauthors{Sanchis-Ojeda et al.~2013}
\shorttitle{Kepler-63b}

\begin{document}

%
\def\ltsima{$\; \buildrel < \over \sim \;$}
\def\lsim{\lower.5ex\hbox{\ltsima}}
\def\gtsima{$\; \buildrel > \over \sim \;$}
\def\gsim{\lower.5ex\hbox{\gtsima}}
%

\bibliographystyle{apj}

\title{
Kepler-63b: A giant planet in a polar orbit around a young Sun-like star
}

\author{
Roberto~Sanchis-Ojeda\altaffilmark{1},
Joshua~N.~Winn\altaffilmark{1},
Geoffrey~W.~Marcy\altaffilmark{2}, 
Andrew~W.~Howard\altaffilmark{3}, 
Howard~Isaacson\altaffilmark{2}, \\
John~Asher~Johnson\altaffilmark{4,5},
Guillermo~Torres\altaffilmark{6}, 
Simon~Albrecht\altaffilmark{1},
Tiago~L.~Campante\altaffilmark{7}, 
William~J.~Chaplin\altaffilmark{7}, \\
Guy~R.~Davies\altaffilmark{7}, 
Mikkel~N.~Lund\altaffilmark{8}, 
Joshua~A.~Carter\altaffilmark{6}, 
Rebekah~I.~Dawson\altaffilmark{6}, 
Lars~A.~Buchhave\altaffilmark{9,10},\\
Mark~E.~Everett\altaffilmark{11,12}, 
Debra~A.~Fischer\altaffilmark{13}, 
John~C.~Geary\altaffilmark{6},
Ronald~L.~Gilliland\altaffilmark{14},
Elliott~P.~Horch\altaffilmark{12,15,16}, 
Steve~B.~Howell\altaffilmark{12,17},
David~W.~Latham\altaffilmark{6}
}

\altaffiltext{1}{Department of Physics, and Kavli Institute for
  Astrophysics and Space Research, Massachusetts Institute of
  Technology, Cambridge, MA 02139, USA}

\altaffiltext{2}{Astronomy Department, University of California,
  Berkeley, CA 94720, USA}

\altaffiltext{3}{Institute for Astronomy, University of Hawaii, 2680 Woodlawn
  Drive, Honolulu, HI 96822, USA}

\altaffiltext{4}{Department of Astronomy, California Institute of Technology, 1200
  E.\ California Blvd., Pasadena, CA 91125, USA}

\altaffiltext{5}{Sloan Fellow; Packard Fellow}

\altaffiltext{6}{Harvard-Smithsonian Center for Astrophysics, Cambridge, MA, 02138, USA}

\altaffiltext{7}{School of Physics and Astronomy, University of Birmingham, Edgbaston, Birmingham B15 2TT, UK}

\altaffiltext{8}{Stellar Astrophysics Centre (SAC), Department of Physics and Astronomy, Aarhus University, Ny Munkegade 120, DK-8000 Aarhus C, Denmark}

\altaffiltext{9}{Niels Bohr Institute, University of Copenhagen, Juliane Maries vej 30, 2100 Copenhagen, Denmark }

\altaffiltext{10}{Centre for Star \& Planet Formation, Natural History Museum of Denmark, University of Copenhagen, {\O}ster Voldgade 5-7, 1350 Copenhagen, Denmark}

\altaffiltext{11}{National Optical Astronomy Observatory, 950 N.\ Cherry Ave, Tucson, AZ 85719}

\altaffiltext{12}{Visiting Astronomer Kitt Peak National Observatory}

\altaffiltext{13}{Astronomy Department, Yale University, New Haven, CT}

\altaffiltext{14}{Center for Exoplanets and Habitable Worlds, The Pennsylvania State University, 525 Davey Lab, University Park, PA 16802 }

\altaffiltext{15}{Southern Connecticut State University, New Haven, CT 06515}

\altaffiltext{16}{Adjunct Astronomer, Lowell Observatory}

\altaffiltext{17}{NASA Ames Research Center, Moffett Field, CA 94035}

 \journalinfo{Draft version}

 \slugcomment{Accepted to the {\it Astrophysical Journal}, 2013 August 04}

\begin{abstract}

  We present the discovery and characterization of a giant planet orbiting the young Sun-like star Kepler-63 (KOI-63, $m_{\rm Kp} = 11.6$, $T_{\rm eff} = 5576$~K, $M_\star = 0.98\, M_\odot$). The planet transits every 9.43 days, with apparent depth variations and brightening anomalies caused by large starspots. The planet's radius is $6.1 \pm 0.2~R_{\earth}$, based on the transit light curve and the estimated stellar parameters. The planet's mass could not be measured with the existing radial-velocity data, due to the high level of stellar activity, but if we assume a circular orbit we can place a rough upper bound of $120~M_{\earth}$ (3$\sigma$). The host star has a high obliquity ($\psi$ = $104^{\circ}$), based on the Rossiter-McLaughlin effect and an analysis of starspot-crossing events. This result is valuable because almost all previous obliquity measurements are for stars with more massive planets and shorter-period orbits. In addition, the polar orbit of the planet combined with an analysis of spot-crossing events reveals a large and persistent polar starspot. Such spots have previously been inferred using Doppler tomography, and predicted in simulations of magnetic activity of young Sun-like stars.

\end{abstract}

  
\section{Introduction}

There are good reasons why planet hunters try to avoid chromospherically active stars. For those who use the radial-velocity technique, starspots and plages distort the absorption lines, inducing radial-velocity signals that can be similar to those of planets (see, for example, Lovis et al.\ 2011). A good example of these complications is CoRoT-7 (L{\'e}ger et al.\ 2009), for which different authors have measured different planet masses based on the same radial-velocity data, due to the strong activity of the host star (Queloz et al.\ 2009, Pont et al.\ 2011, Ferraz-Mello et al.\ 2011, Hatzes et al.\ 2011).

Starspots can also be a source of noise in the transit technique. Starspots are carried around the star by rotation, inducing flux variations that could be hard to detect from ground-based telescopes. When they go unnoticed, they can bias the determination of the transit parameters (Czesla et al.\ 2009, Carter et al.\ 2011). In addition, when the planet crosses over a dark starspot, it temporarily blocks less light than expected, causing a brightening anomaly (Silva 2003). These can be an additional source of error, or be confused with transits of other bodies in the system (Pont et al.\ 2007, Rabus et al.\ 2009).

Space-based transit surveys have the potential to overcome these problems, thanks to their high photometric precision and nearly continuous time coverage. The data from these surveys provide the opportunity to study the general activity levels of thousands of stars (Basri et al.\ 2011) as well as spot evolution and magnetic cycles of individual systems (Bonomo \& Lanza 2012). With hundreds or even thousands of transiting objects detected to date, spot-crossing events are more readily observed. They bear information about the sizes, temperatures, and positions of the spots (Silva 2003), as well as the stellar rotation period (Silva-Valio 2008; Dittmann et al.\ 2009).

Spot-crossing events can also provide information about the architecture of exoplanetary systems. Measurements of the angle between the spin axis of the star and the orbital plane of the planet (known as the obliquity) can help test theories of formation and evolution of these systems (Queloz et al.\ 2000, Winn et al.\ 2005). Most of the obliquity measurements to date have been based on the Rossiter-McLaughlin (RM) effect, a spectroscopic effect observed during transits (see, e.g., the recent compilation by Albrecht et al.\ 2012). One can also test whether a transit-hosting star has a high obliquity using asteroseismology (Gizon \& Solanki 2003, Chaplin et al.\ 2013), the combination of $v \sin i_\star$, stellar radius, and stellar rotation period (Schlaufman 2010; Hirano et al.\ 2012), or starspot-crossing events (Sanchis-Ojeda et al.\ 2011, Nutzman et al.\ 2011, D\'{e}sert et al.\ 2011, Tregloan-Reed et al.\ 2013).

The basic idea behind using starspot-crossing events to measure the obliquity is that when the obliquity is low, any such events are expected to recur in consecutive transits. This is because in such cases the trajectory of the spot is parallel to the trajectory of the planet across the stellar disk; when the planet transits again, the spot is likely to have remained on the transit chord and a spot-crossing event will occur at a later phase of the transit. In contrast, the rotation of a highly oblique star would carry the spot away from the transit chord, and the anomalies would not recur in consecutive transits. For a more detailed explanation and recent elaborations of this technique, we refer the reader to Sanchis-Ojeda et al.\ (2013). One interesting feature of the starspot-crossing technique is that in the case of highly misaligned system, the planet may transit across a wide range of stellar latitudes (Deming et al.\ 2011; Sanchis-Ojeda \& Winn 2011). In these systems we have the rare opportunity to measure the latitudes of starspots and their evolution in time.

In this paper we present Kepler-63b, a new transiting planet discovered with the \textit{Kepler} space telescope. The paper is organized as follows. Section~\ref{sec:obs} describes the \textit{Kepler} observations and other follow-up observations necessary to confirm the planetary nature of Kepler-63b. Section~\ref{sec:stellarcharac} describes the effort to characterize the age, radius and mass of the Kepler-63 host star. Section~\ref{sec:planet} explains how we characterized the planet in the presence of large and dark starspots. Section~\ref{sec:RM} demonstrates that Kepler-63 has a large obliquity, using the RM effect. Section~\ref{sec:obliquity} confirms the high obliquity of the system using spot-crossing events. Section~\ref{sec:magnetic} summarizes what we have learned about the starspots on Kepler-63, including their latitudes. The paper finishes with a discussion of the results in the context of current theories for stellar activity and planetary systems.

\section{Observations}
\label{sec:obs}

\subsection{\textit{Kepler} observations}
\label{sec:kep}

For more than 4 years, the \textit{Kepler} space telescope monitored approximately 150,000 stars in the constellations of Cygnus and Lyra (Koch et al.\ 2009; Borucki et al.\ 2010). The observations consisted of a series of 6~s exposures that were combined into final images with an effective exposure time of 1~min (short-cadence mode, Gilliland et al.\ 2010) or 29.4~min (long-cadence mode). The target star Kepler-63 was identified in the \textit{Kepler} Input Catalog (Brown et al.\ 2011) as KIC~11554435 (also 2MASS~J19165428+4932535) with \textit{Kepler} magnitude 11.58 and $V = 12.02$. Because of the relative brightness of the star, and the high signal-to-noise ratio (S/N) of the flux dips, it was identified as a transiting-planet candidate early in the mission and designated Kepler Object of Interest (KOI) number 63. For this study we used long-cadence observations from quarter 1 through 12, spanning nearly three years (2009~May~13 through 2012~March 28). Short-cadence observations were also used whenever available (quarters 3-12, as well as one month in quarter 2).

The \textit{Kepler} pipeline provides data with time stamps expressed in barycentric Julian days in the TDB (Barycentric Dynamical Time) system. Two sets of fluxes are provided: simple aperture photometry, which is known to be affected by several instrumental artifacts (Jenkins et al.\ 2010); and fluxes that have been corrected with an algorithm called PDC-MAP that attempts to remove the artifacts while preserving astrophysical sources of variability (Stumpe et al.\ 2012; Smith et al.\ 2012). For this study we used the PDC-MAP time series. Because of the large pixel scale (4 arcseconds) of {\it Kepler}'s detectors it is always important to consider the possibility that the reported fluxes include contributions from neighboring stars (``blends''). The time series of the measured coordinates of the source of light are useful diagnostics. For Kepler-63, we used the coordinates of the center-of-light based on PSF-fitting (PSF\_CENTR) which were provided by the {\it Kepler} pipeline (see~\ref{sec:back}).

\subsection{Spectroscopic observations}
\label{sec:spectroscopy}

We used the spectroscopic observations gathered by the {\it Kepler} Follow-up Program (KFOP). One spectrum was obtained with fiber-fed Tillinghast Reflector {\'E}chelle Spectrograph (TRES) on the 1.5m Tillinghast Reflector at the Fred Lawrence Whipple Observatory on Mt.\ Hopkins, Arizona, with a resolution of $44,000$. The observation took place on 2009~June~13 with a exposure time of 24~min, giving a S/N of 64 in the Mg~I~b order. Another spectrum was taken with the HIRES spectrograph (Vogt et al.\ 1994) on the 10m Keck~I telescope at Mauna Kea, Hawaii, with a resolution of $48,000$. The observation took place on 2009~August~1 with an exposure time of 20~min, giving a S/N of 250. Three more spectra were taken with the FIber-fed {\'E}chelle Spectrograph (FIES) on the 2.5m Nordic Optical Telescope (NOT) on La Palma, Spain (Djupvik \& Andersen 2010), with a resolution of $46,000$. The observations took place on 2010 June 2, 5 and 6, with a typical exposure time of 20~min, giving a S/N of about 80.

We conducted additional observations to try to measure the radial-velocity signal induced by the planet on the star, and also to detect the RM effect. We used HIRES on the Keck~I 10m telescope to obtain 7 spectra during the two weeks before the transit of 2011~August~20/21. Then, during the night of the transit, we obtained 30 spectra with a typical exposure time of 10~minutes, starting 3~hours before the transit and finishing 3~hours afterward. We determined relative radial velocities in the usual way for HIRES, by analyzing the stellar spectra filtered through an iodine cell (wavelength range 500--600~nm). For the analysis, we used a modified version of the original code by Butler et al.\ (1996).  Table~\ref{tbl:rv} gives the radial velocities.

\begin{deluxetable}{lcc}

\tabletypesize{\scriptsize}
\tablecaption{Relative Radial Velocity Measurements of Kepler-63\label{tbl:rv}}
\tablewidth{0pt}

\tablehead{
\colhead{BJD$_{\rm TDB}$} &
\colhead{RV [m~s$^{-1}$]} &
\colhead{Unc.~[m~s$^{-1}$]}
}

\startdata
$ 2455782.054444  $  &   $ -19.6   $   &   $ 3.3 $ \\
$ 2455782.937924  $  &   $  31.3   $   &   $ 3.0 $ \\
$ 2455787.771381  $  &   $ -26.7   $   &   $ 3.0 $ \\
$ 2455788.809546  $  &   $  -5.6   $   &   $ 3.2 $ \\
$ 2455789.858820  $  &   $  -7.5   $   &   $ 3.1 $ \\
$ 2455790.807157  $  &   $ -23.2   $   &   $ 2.8 $ \\
$ 2455792.775137  $  &   $ -37.4   $   &   $ 2.7 $ \\
$ 2455793.743264  $  &   $   4.2   $   &   $ 2.2 $ \\
$ 2455793.750173  $  &   $  -4.9   $   &   $ 2.1 $ \\
$ 2455793.757280  $  &   $   0.6   $   &   $ 2.1 $ \\
$ 2455793.791469  $  &   $  -2.5   $   &   $ 2.2 $ \\
$ 2455793.798830  $  &   $   1.3   $   &   $ 2.2 $ \\
$ 2455793.806157  $  &   $  -2.7   $   &   $ 2.2 $ \\
$ 2455793.813541  $  &   $  -2.3   $   &   $ 2.0 $ \\
$ 2455793.821018  $  &   $  -1.4   $   &   $ 2.3 $ \\
$ 2455793.828321  $  &   $   7.8   $   &   $ 2.2 $ \\
$ 2455793.835717  $  &   $   2.2   $   &   $ 2.1 $ \\
$ 2455793.843124  $  &   $   2.7   $   &   $ 2.2 $ \\
$ 2455793.850578  $  &   $  13.2   $   &   $ 2.1 $ \\
$ 2455793.857834  $  &   $  16.6   $   &   $ 2.6 $ \\
$ 2455793.865369  $  &   $  11.6   $   &   $ 2.4 $ \\
$ 2455793.872800  $  &   $  18.8   $   &   $ 2.4 $ \\
$ 2455793.880404  $  &   $  20.8   $   &   $ 2.4 $ \\
$ 2455793.887626  $  &   $   9.9   $   &   $ 2.3 $ \\
$ 2455793.895195  $  &   $   7.3   $   &   $ 2.2 $ \\
$ 2455793.902278  $  &   $  15.1   $   &   $ 2.8 $ \\
$ 2455793.912105  $  &   $   0.3   $   &   $ 3.0 $ \\
$ 2455793.919987  $  &   $  12.5   $   &   $ 3.3 $ \\
$ 2455793.927382  $  &   $   5.9   $   &   $ 2.6 $ \\
$ 2455793.934894  $  &   $  -7.0   $   &   $ 2.7 $ \\
$ 2455793.942232  $  &   $   3.8   $   &   $ 2.8 $ \\
$ 2455793.949546  $  &   $  -7.3   $   &   $ 2.9 $ \\
$ 2455793.957266  $  &   $  -3.3   $   &   $ 3.3 $ \\
$ 2455793.964546  $  &   $ -19.0   $   &   $ 3.1 $ \\
$ 2455793.972081  $  &   $   0.7   $   &   $ 3.1 $ \\
$ 2455793.979569  $  &   $  -6.7   $   &   $ 3.1 $ \\
$ 2455794.010807  $  &   $ -10.1   $   &   $ 2.9 $
\enddata

\tablecomments{RVs were measured relative to an arbitrary template
  spectrum; only the differences are significant. Column 3 gives the internally estimated measurement uncertainty
  which does not account for any ``stellar jitter.''}

\end{deluxetable}

\normalsize

\subsection{Speckle imaging}
\label{sec:speckle}

High-resolution images are useful to establish which stars are contributing to the {\it Kepler} photometric signal. Speckle imaging was conducted on the night of 2010~September~17, using the two-color DSSI speckle camera at the WIYN 3.5m telescope on Kitt Peak, Arizona. The speckle camera simultaneously obtained images in two filters: $V$ (5460/400\AA) and $R$ (6920/400\AA). These data were processed to produce a final reconstructed speckle image for each filter. Details of the processing were presented by Howell et al.\ (2011). The speckle observations are sensitive to companions between 0.05--1.5 arcseconds from Kepler-63. We found no companion star within this range of separations, and can place an upper limit on the brightness of such stars corresponding to $\Delta R =$~4-5~mag and $\Delta V=$~3-5~mag below the brightness of Kepler-63.

\section{Stellar characterization}
\label{sec:stellarcharac}

\subsection{Rotation period and age estimate}
\label{sec:gyro}

We identified Kepler-63 as an interesting target based on its high level of chromospheric activity, and the high S/N with which individual transits are detected. The high level of activity is evident from the quasi-periodic stellar flux variations. To study these variations we used the published transit ephemeris (Batalha et al.~2013) to remove the transit signals. To reject outliers we clipped those data points more than 3$\sigma$ away from the median flux over the surrounding 10~hr interval. Since the star's light fell on a different CCD during each of the four {\it Kepler} observing seasons, we had to make a choice for the normalization of each quarterly time series. We chose to divide each quarterly time series by the mean flux in that quarter, since it seemed to be the easiest way to avoid large flux discontinuities between quarters.

\begin{figure*}[ht]
\begin{center}
\leavevmode
\hbox{
\epsfxsize=5.2in
\epsffile{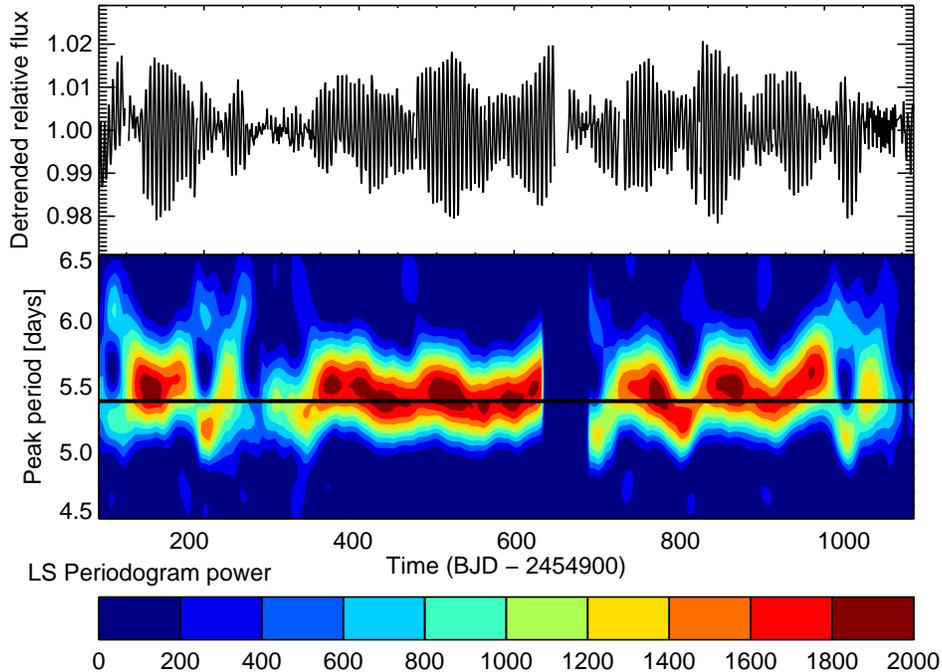}}
\end{center}
\vspace{-0.1in}
\caption{ {\bf Determination of the stellar rotation period from {\it Kepler} photometry} {\it Top.}---The time series used to estimate the rotation period, after removing the transit signals. The short-term variability is presumably
from spots being carried by the stellar rotation. The longer-term variations
may be caused by spot evolution. {\it Bottom.}---Running periodogram. Red colors represent high power, and the black line is where $P_{\rm rot}/P_{\rm orb} = 4/7$.}
\label{fig:activity}
\vspace{0.1in}
\end{figure*}

The upper panel of Figure~\ref{fig:activity} shows the relative flux time series. The short-term variability on the scale of a few days is presumably caused by spots being carried around by the rotation of the star. There is also long-term variability, probably reflecting spot evolution. The peak-to-peak variability reaches a maximum value near $4\%$. There are also intervals with much lower variability. A Lomb-Scargle periodogram (Scargle 1982) shows a strong peak at 5.4~days, with a full width at half-maximum of 0.014~days. Sometimes the highest peak in a periodogram actually represents a harmonic of the true period, but in this case there is no significant signal at twice the candidate period and there is a (weaker) signal at half the candidate period, supporting our identification. Therefore we interpret the strongest peak as the rotation period, and adopt $P_{\rm rot} = 5.401 \pm 0.014$ days.

This is nearly commensurate with the orbital period of Kepler-63b, with $P_{\rm rot}/P_{\rm orb} = (4.01 \pm 0.01)/7$.  Whether this relationship is caused by a physical process or is merely a coincidence, it has an important consequence for the interpretation of spot-crossing anomalies, as pointed out by Winn et al.\ (2010b) in the context of the HAT-P-11 system. The near-commensurability causes a ``stroboscopic'' effect in the pattern of anomalies, and may lead to the detection of recurrences even in the case of misaligned systems. For example, D{\' e}sert et al.\ (2011) used the stroboscopic effect to boost the S/N of the spot signals in the Kepler-17b system.

Other lower-power peaks are present in the periodogram, which could be a sign of differential rotation or spot evolution. We checked for any variability of the position of the highest-power peak by computing a running periodogram: for each time $t$ we calculated the Lomb-Scargle periodogram of a 50-day time interval centered on $t$. We performed these periodograms for a sequence of $t$ values spaced apart by 5~days. Since there are occasional gaps in the data collection, we only computed periodograms for those intervals for which the gaps constitute less than 20\% of the interval. (Removing the transits only eliminates 2\% of the data.) The results are shown in the lower part of Figure~\ref{fig:activity}, with red colors indicating higher periodogram power. The black line is where $P_{\rm rot}/P_{\rm orb} = 4/7$. It seems that the highest peak was quite stable over the three years of observations.

Our spectroscopic analysis (see section \ref{sec:spectroscopy}), shows that Kepler-63 is a Sun-like star. With such a short rotation period of 5.4 days, and assuming the star has not been spun up due to tidal or other interactions, Kepler-63 is likely to be relatively young. The Sun had a rotation period of 6 days at an age of approximately 300~Myr, based on the Skumanich (1972) law in which the rotation period grows as the square root of time. None of the Sun-like stars in the Hyades or Praesepe have rotation periods as short as 5.4 days (Irwin \& Bouvier 2009). Thus Kepler-63 is likely younger than 650~Myr, the approximate age of these clusters. Schlaufman~(2010) presented a convenient polynomial relationship between stellar mass, age, and rotation period; using this relationship we find for Kepler-63 an age of $210\pm 35$~Myr. We also used the polynomial relationship given by Barnes (2007) and updated by Meibom et al.\ (2008); applying this relationship with $B-V = 0.7$ for Kepler-63 gives an age of $210\pm 45$~Myr.

Independent evidence for youth comes from a high observed level of chromospheric emission. The Mt.\ Wilson $S_{\rm HK}$ index, obtained from the Keck spectra, has an average value of 0.37, from which we estimate a chromospheric flux ratio $\log{R'_{\rm HK}} = -4.39$. Using the correlation between chromospheric emission and rotation period presented by Noyes et al.\ (1984), and an estimated $B-V = 0.7$ for this star, one would expect the star's rotation period to be 6.8~days, in good agreement with our photometrically-derived period. The relationship presented by Mamajek \& Hillenbrand (2008) between $\log{R'_{\rm HK}}$ and stellar age gives in this case an age of 280~Myr.

\subsection{Stellar dimensions}
\label{sec:stellardim}

At first we determined the star's spectroscopic parameters by applying the spectral synthesis code Spectroscopy Made Easy (SME; Valenti \& Fischer~2005) to the Keck spectrum. The results were $T_{\rm eff} = 5698 \pm 44$\,K, ${\rm [Fe/H]} = +0.26 \pm 0.04$, $v \sin i_\star = 5.8 \pm 0.5~$km~s$^{-1}$, and $\log g = 4.64 \pm 0.06$. We then recognized that this value of $\log\,g$ is anomalously high for a young Sun-like star, which was otherwise suggested by the star's effective temperature and relatively short rotation period. One would instead expect $\log\,g$ to be closer to the Solar value of 4.44.  In fact, by consulting the Yonsei-Yale (Y$^2$) stellar evolution models \citep{Yi:01} we found that the SME-based value of $\log g$ would place the star in a very unusual location on the theoretical H-R diagram, at a higher gravity than any of the isochrones for the nominal metallicity. This mismatch probably reflects the well-known biases in spectroscopic determinations of $\log g$ (see, e.g., Torres el al.\ 2012).

To address this issue we used an updated version of the Stellar Parameter Classification code \citep[SPC;][]{Buchhave:12} which determines the spectroscopic parameters subject to a prior constraint on the surface gravity. In this implementation, SPC is used to provide an initial guess at the effective temperature and metallicity for the star and select the Y$^2$ evolutionary models which are compatible with this initial guess within fairly wide intervals ($\pm$250~K in $T_{\rm eff}$, $\pm$0.3~dex in [m/H]\footnote{Throughout this paper the generic metallicity index [m/H] computed with SPC in the \ion{Mg}{1}\,b region will be considered as equivalent to the more commonly used [Fe/H] index, as is usually the case for stars with near-solar composition.}). The selected evolutionary models then provide an interval of allowed surface gravities, which are used to construct a prior on the surface gravity for a second iteration of SPC. The final results, based on weighted averages of the results for all of the spectra (3 FIES, 2 TRES, and one HIRES), are: $T_{\rm eff} =5576 \pm 50~K$, ${\rm [m/H]} = 0.05 \pm 0.08$, $\log g = 4.52 \pm 0.10$, and $v\sin i_\star = 5.4 \pm 0.5~\kms$.

Finally, we used the Y$^2$ models to determine the stellar dimensions, based not only on the spectroscopic parameters but also the rotation-based constraint on the stellar age (see section~\ref{sec:gyro}). A comparison of our spectroscopic parameters with the Y$^2$ models for ages between 100 and 600 Myr suggests a surface gravity of $\log g = 4.52 \pm 0.02$. Using the initial spectroscopic parameters, but with the smaller 0.02 dex uncertainty on $\log g$, we then used the Y$^2$ models to determine the stellar mass ($M_{\star}$), radius ($R_{\star}$), and mean density ($\rho_{\star}$). We proceeded in a Monte Carlo fashion, randomly drawing 100,000 sets of temperature, metallicity, and surface gravity values from assumed Gaussian distributions of those quantities based on the above results, and inferring the stellar properties for each set. Our final results were obtained from the mode of the corresponding posterior distributions, and the (``1-$\sigma$'') uncertainties from the 15.85\% and 84.15\% percentiles of the cumulative distributions. These values are reported in Table~\ref{tbl:params}. Based on the absolute and apparent magnitudes, the distance to the system is $200 \pm 15$~pc.

\section{Planet characterization}
\label{sec:planet}

\subsection{Constraints on blend scenarios}
\label{sec:back}

The speckle image of Kepler-63 (see section \ref{sec:speckle}) puts tight constraints on any background star that could be responsible for the transit signal. We also used the PSF-fitted image centroids provided with the {\it Kepler} data to further restrict the possibilities for background blend scenarios (Batalha et al.\ 2010; Bryson et al.\ 2013). Using a similar approach as Chaplin et al.\ (2013), we selected the long-cadence column and row centroids within a 2~hr interval centered on each transit, and used the surrounding three hours of data before and after the transits to correct for linear trends caused by pointing drifts and other instrumental effects. We phase-folded the centroid data and computed the mean differences between the row and column values inside and outside of the transits. The centroid shifts were $ 12 \pm 11 \,\mu$pix in the column direction and $28 \pm 14\, \mu$pix in the row direction.

If the source of the transits were a background star situated at a distance $\Delta x$ from Kepler-63, the expected centroid shift would be approximately $dx = (\Delta x) \, \delta$, where $\delta$ is the transit depth. Adding both the row and column shift in quadrature, we obtain $dx = 30 \pm 18 \, \mu$pix, or $dx < 84 \, \mu$pix (3$\sigma$). The radius of confusion $r$, defined as the maximum angular separation of any hypothetical background source that could be responsible for the transit signal, can be obtained dividing by the transit depth and multiplying by 4~arcsec~pix$^{-1}$, giving $r = 0.084$~arcsec. The more sophisticated techniques described by Bryson et al.~(2013) indicate that the source of photometric variability is offset from the bright target star by $0.02 \pm 0.02$~arcsec (based on the Data Validation Reports provided by the \textit{Kepler} team). These results are compatible with Kepler-63 as the origin of the flux variations, and require any hypothetical background source of the variations to be aligned with Kepler-63 to within a small fraction of an arcsecond, an unlikely coincidence.

This type of analysis cannot exclude the possibility that the transit signal is caused by a planet orbiting a companion star that is gravitationally bound to the intended target star. However, the starspot anomalies that are detected in many transits show that the planet is orbiting a heavily spotted star. This star must be the main source of light in the aperture, because the $4\%$ flux variations observed would be unphysically large if they actually represented the diluted variations of a fainter star. This possibility is also excluded by the good agreement between the temperature of the occulted spots and the size of the flux variations observed (see section \ref{sec:magnetic}). We conclude that the transit-like signals do indeed arise from transits of a planet around the star Kepler-63.

\subsection{Transit analysis}
\label{sec:trans}

\begin{figure}[ht]
\begin{center}
\leavevmode
\hbox{
\epsfxsize=3.5in
\epsffile{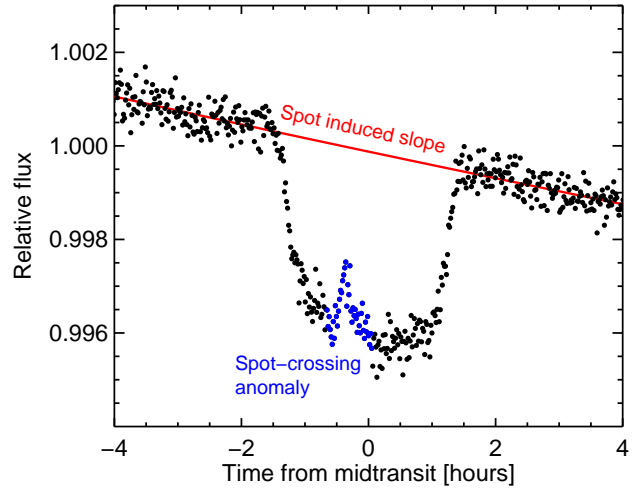}}
\end{center}
\vspace{-0.1in}
\caption{ {\bf Example of the effects of starspots on transit signals.} The black dots are {\it Kepler} data points
for a particular transit. Spot-crossing anomalies (blue dots) are identified
visually and masked out. Variability on the longer timescale is modeled as a second-order polynomial (red line). }
\label{fig:transit}
\vspace{0.1in}
\end{figure}

To obtain accurate transit parameters we needed to correct for the effects induced by starspots (see, e.g., Czesla et al.\ 2009; Carter et al.\ 2011; Sanchis-Ojeda et al.\ 2012). We chose to work only with the short-cadence data, since the 30-minute time sampling of long-cadence data is too coarse to allow a clear identification of the spot-crossing events. We defined the transit window as an interval of 4~hr centered on the expected transit time. The out-of-transit (OOT) portion was defined as the 2~hr preceding the transit window plus the 2~hr following the transit window, giving a total of 4~hr of data. First, the data from each transit window were normalized such that the OOT data had a median flux of unity. Figure~\ref{fig:transit} shows an illustrative example. We visually inspected the 96 transit light curves and identified 145 spot-crossing events. All of them were temporary brightenings (rather than fadings), implying that the detectable spots on the surface of Kepler-63 have a temperature lower than the photosphere. This is in accordance with a general trend observed for very active stars (Foukal 1998). In order to properly estimate the transit parameters, the data points within these anomalies were assigned zero weight in the fits. More than $25\%$ of the in-transit data points were assigned zero weight, which speaks to the high level of activity on the particular region that the planet is transiting.

Figure~\ref{fig:transit} also shows the effect of the longer-term variations in stellar flux.  Large starspots combined with the short rotation period introduce strong gradients in the out-of-transit flux, and occasionally significant curvature, especially if the transit happens near a flux minimum or maximum. As long as we have removed the spot-anomalies correctly, the observed flux can be described as (cf.\ Carter et al.\ 2011)
\begin{equation}
\label{eq:flux-with-spots}
F(t) = F_{0}\left[ 1-\epsilon (t)\right] - \Delta F(t)
\end{equation}
where $F_0$ is the stellar flux in the absence of transits or spots,
$\epsilon (t)$ is the fractional loss of light due to starspots, and $\Delta F(t)$ is the flux blocked by the planet. The function $\epsilon (t)$ is responsible for the overall
gradient in time throughout the transit (see the red line in Figure \ref{fig:transit}). To obtain
the normalized flux we divide
Eqn.~(\ref{eq:flux-with-spots})
by $F_0[1-\epsilon(t)]$. We then model the OOT variation by a second-order
polynomial in time, giving
\begin{equation} \label{eq:flux}
f(t)  \approx 1 + c_0 + c_1(t-t_{c}) + c_2(t-t_{c})^2  - \frac{1}{1-\epsilon(t_c)}\frac{\Delta F(t)}{F_0},
\end{equation}
where the out-of-transit variation is described by a second-order polynomial in time, and $\epsilon(t_c)$ represents the relative flux lost due to spots at the time of transit $t_c$. To determine the coefficients $c_i$, we fitted a second-order polynomial to the OOT portion of the data. We then subtracted the best-fitting polynomial from the data in the entire transit window, to ``rectify'' the data. The loss of light due to the planet, $\Delta F(t)/F_0$, was assumed to be the same for all transits, but each transit was assigned an independent value of $\epsilon(t_c)$. To avoid having to fit all the data with hundreds of parameters, we performed the transit modeling in four stages, described below.

\subsubsection{Step 1. Initial folded light curve analysis.} 

First we needed good initial guesses for the transit parameters. We created a phase-folded light curve based on the normalized transits, using the orbital period from Batalha et al.\ (2013). We averaged the phase-folded light curve into 4~min bins, chosen to improve computation speed without a significant loss of accuracy. At this stage we ignored the transit-to-transit variations and simply modeled the folded light curve with an idealized Mandel \& Agol (2002) model, the free parameters being $(R_p/R_s)^2$, $R_s/a$, the impact parameter $b$, and two linear combinations of the quadratic limb-darkening coefficients (chosen to minimize correlations as recommended by P{\'a}l 2008).

\subsubsection{Step 2. Individual transit analysis.} 

Next we wanted to obtain individual transit times and depths.  The 5 parameters of the best-fitting model from step 1 were held fixed, and the data from each transit window were fitted with three additional parameters: the time of transit $t_c$, the linear coefficient $c_1$ from Eqn.~(\ref{eq:flux}), and the spot-coverage factor $\epsilon(t_c)$ (which for brevity we hereafter denote simply $\epsilon$). The linear coefficient $c_1$ was allowed to vary because it is covariant with the transit time.  We assigned an uncertainty of 279~ppm to each individual SC data point, as this is the standard deviation of the OOT portion of the unbinned folded light curve. We found the best-fitting model for each individual transit light curve and used a Markov Chain Monte Carlo (MCMC) algorithm to explore the allowed parameter space.

\begin{figure*}[ht]
\begin{center}
\leavevmode
\hbox{
\epsfxsize=5.0in
\epsffile{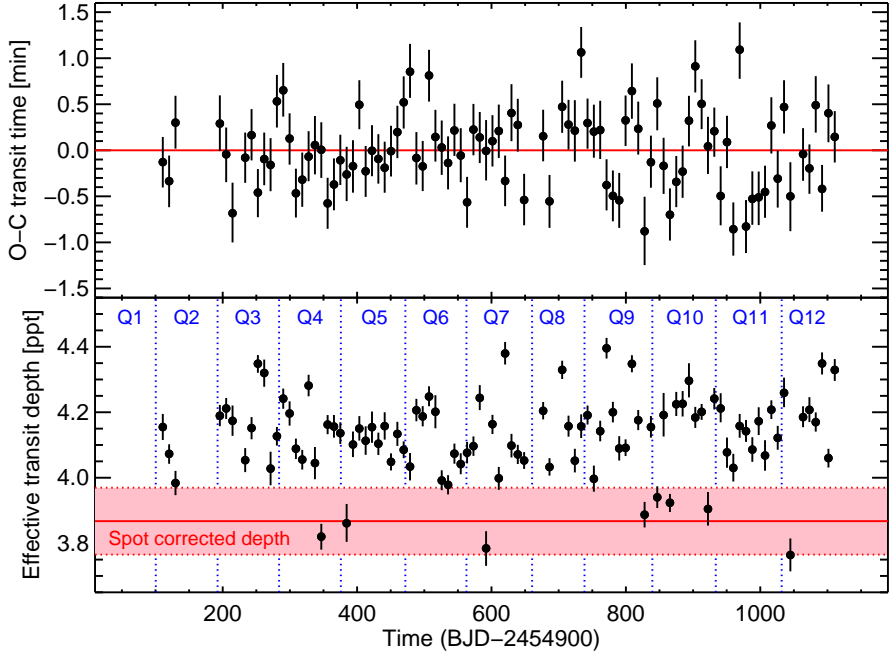}}
\end{center}
\vspace{-0.1in}
\caption{ {\bf Variations in transit times and apparent depths.} {\it Top.}---Residuals of a linear fit to the transit times of Kepler-63b. The excess of scatter is significant ($\chi^2 = 202$ with 94 degrees of freedom) and is likely due to uncorrected spot-crossing events. {\it Bottom.}---The black dots represent the effective depth of individual transits in parts per thousand (ppt). No strong correlation with the quarter is observed, suggesting that the variability is mostly due to starspots as opposed to variable amounts of blended flux. The final adopted value for the transit depth (red line) and its 1$\sigma$ confidence interval (shaded region) were obtained by assuming that the star is nearly spot-free during the transits with the smallest effective depth (see section~\ref{sec:trans}).}
\label{fig:ttv}
\vspace{0.1in}
\end{figure*}

We fitted a linear function of epoch to the transit times and used it to estimate the orbital period and a particular transit time (chosen to be the first transit observed with short cadence). Figure~\ref{fig:ttv} shows the residuals between the observed and calculated transit times. There is no clear structure, but the fit has $\chi^2=202$ with 94 degrees of freedom, suggesting that the uncertainties on the transit times have been underestimated. We attribute this excess scatter to uncorrected effects of stellar spots (see, e.g., Sanchis-Ojeda et al.\ 2011, Oshagh et al.\ 2013b), although transit-timing variations could also be present due to another planet orbiting the same star (Agol et al.\ 2005; Holman \& Murray 2005; Nesvorn{\'y} et al.\ 2012). To account for the excess scatter we enlarged the uncertainties in the orbital period and the transit epoch by 46\% (such that $\chi^2 = N_{\rm dof}$).

In our procedure each transit is associated with a particular value of $\epsilon$, but it is a more common practice to report an effective depth for each transit. The transit depth obtained in this way would be equivalent to the transit depth fixed in step 1, shared by all transits, divided by $1-\epsilon$. In the lower panel of Figure~\ref{fig:ttv}, we plot this effective depth for each transit. The variability of the apparent transit depths is as high as 10\%. There is no clear correlation of the apparent depths with the {\it Kepler} quarter, implying that contamination from background stars, if present, must be very small or common to all the photometric apertures used in the different quarters. The observed variability is most likely induced by starspots.

\subsubsection{Step 3. Choice of the baseline transit depth.} 

In order to obtain the final transit parameters, we readjusted the scale by which $\epsilon$ is measured, thereby renormalizing the transits. The distribution of spot coverage factors ($\epsilon$) had a mean near zero and standard deviation close to $3\%$. This choice would make sense if the effects of dark starspots and bright plages were comparable on average. However, we did not detect any plage-crossing events whereas we did detect many spot-crossing events. Assuming that dark starspots dominate the stellar flux variations, the true loss of light should always be positive, and the shallowest effective transit depth should occur when the star has the smallest spot coverage. For that particular transit the true $\epsilon \approx 0$ and the rest of spot coverage factors are positive (Carter et al.\ 2011, Sanchis-Ojeda et al.\ 2012).

A simple approach would be to identify the transit with the smallest value of $\epsilon$, and subtract that value from the entire distribution of $\epsilon$ values. We chose instead a more robust method that does not depend entirely on a single $\epsilon$ value. First we removed outliers from the distribution of $\epsilon$ factors using a 3$\sigma$ clipping algorithm. Then we assumed that the distribution is Gaussian and computed the standard deviation of the remaining $\epsilon$ factors. Finally we shifted the distribution of $\epsilon$ values to force the median of the distribution to be two standard deviations above zero. This procedure ensures that most of the $\epsilon$ factors are positive.

\subsubsection{Step 4. Final transit parameters.} 

To obtain the final transit parameters, we used the $\epsilon$ values determined in step 3 to renormalize each transit light curve.  The intention was to correct all of the transit signals to have a uniform shape, similar to what one would observe if the planet were transiting a spot-free star. The transit data were then folded using the newly calculated linear ephemeris, binned to have a time sampling of one minute, and modeled with the same five-parameter transit model used in step 1.  Table~\ref{tbl:params} gives the final parameter values, with uncertainties estimated using an MCMC algorithm. The uncertainty in the transit depth was enlarged beyond the statistical uncertainty, to take into account the procedure we have described in step 3. We used the width of the distribution of measured depths as the measure of systematic uncertainty in the transit depth. The final value of the depth with the enlarged uncertainty is depicted in Figure \ref{fig:ttv}.

\subsection{Orbital eccentricity}
\label{sec:ecc}

The orbital period of Kepler-63b is long enough that it is not safe to assume the orbital eccentricity has been damped by tides to a negligible level. We attempted to learn the orbital eccentricity in two different ways: by searching for occultations (secondary eclipses); and by using the mathematical relationship between the transit parameters, the mean stellar density, and the orbital eccentricity and argument of pericenter.

If the orbit were circular, occultations would occur halfway in between transits, with a duration equal to that of the transits.  For an eccentric orbit, the timing and duration of the secondary eclipse depend on the eccentricity $e$ and the argument of periastron $\omega$ (see, e.g., Winn 2011). A grid search was performed to detect the occultation, but the result was negative. This is not surprising, since the occultation depth would be of order $(R_p/a)^2$ = 10 ppm, below our level of detectability.

One can also obtain information about the eccentricity of the orbit by combining the orbital period, scaled semimajor axis, and mean stellar density $\rho_{\star}$ (see, e.g.,
Dawson \& Johnson 2012). We compute
\begin{equation}
\frac{1+e\sin{\omega}}{\sqrt{1-e^2}} \approx \frac{2 \sqrt{R_p/R_\star}}{\sqrt{T^2_{14}-T^2_{23}}}\left(\frac{3P}{G\pi^2 \rho_{\star}}\right)^{1/3} = 0.92 \pm 0.02,
\end{equation}
where $T_{14}$ is the duration between first and fourth contact, and $T_{23}$ is the duration between second and third contact. This expression is close to unity when the eccentricity is low.  Based on the measured values of $T_{14}$, $T_{23}$, $P$, and $R_p/R_\star$, and the estimated value of $\rho_\star$ from section~\ref{sec:stellardim} we find $e<0.45$ (3$\sigma$).  The formal 68.3\% confidence interval is 0.08--0.27.  Thus the eccentricity is not likely to be very high, but moderate values cannot be excluded.

\begin{deluxetable*}{lccc}
\tabletypesize{\scriptsize}
\tablecaption{System Parameters of Kepler-63\label{tbl:params}}
\tablewidth{0pt}

\tablehead{
\colhead{Parameter} & \colhead{Value} & \colhead{68.3\% Conf.~Limits} & \colhead{Note}
}

\startdata
KIC number / KOI number                                            &  11554435 / 63 & \\
{\it Kepler} apparent magnitude                                    &  11.582 & \\
Right ascension (J2000)                                            &  19$^{\rm h}$16$^{\rm m}$54\fs 28  &   &  \\
Declination (J2000)                                                &  $+49^\circ 32\arcmin 53\farcs52$  &   & \\
 & & \\
Stellar surface gravity, log($g$~[cm~s$^{-2}$])                             & 4.52        & $\pm 0.02$         &  a  \\   
Stellar effective temperature, T$_{\rm eff}$~[K]                            &   5576       & $\pm 50$           & b  \\  
Stellar metallicity [Fe/H]                                                 & 0.05         & $\pm 0.08$         &   b  \\
Stellar mass, M$_\star$~[$M_{\odot}$]                                & $0.984$     & $-0.04$, $+0.035$  & b \\
Stellar radius, $R_\star$~[$R_{\odot}$]                              &  $0.901$  & $-0.022$, $ +0.027$  & b  \\
Stellar mean density, $\rho_\star$~[$\rho_{\odot}$]                  &  $1.345$  & $-0.083$, $+0.089$   & b  \\
Stellar luminosity, $L_\star$~[$L_{\odot}$]                          &  $0.696$  & $-0.059$, $ +0.076$  & b  \\
Stellar rotation period~[days]                                     & 5.401     & $\pm 0.014$          & c   \\   
Mt.\ Wilson chromospheric index $S_{\rm HK}$                        & 0.37 &                           &  d  \\ 
Chromospheric flux ratio $\log{ R'_{\rm HK}}$                        & $-4.39$ &                         &   d  \\  
Distance from Earth~[pc]                                           & 200       & $\pm 15$             &  b  \\ 
 & & \\
Reference epoch~[BJD$_{\rm TDB}$]                                    & $2455010.84307$  &  $\pm 0.00005$  &  e   \\
Orbital period~[days]                                               & $9.4341505$      &  $\pm 0.0000010$  & e  \\
Planet-to-star radius ratio, $R_{\textrm{p}}/R_\star$                 & $0.0622$         &  $\pm 0.0010$  &   e   \\
Transit impact parameter, $b$                                       & $0.732$           &  $\pm 0.003$  &   e   \\
Scaled semimajor axis, $a/R_\star$                                   & $19.12$           &  $\pm 0.08$   &  e  \\
Transit duration (1st to 4th contact)~[hr]                           & $2.903$          &  $\pm 0.003$  &   e  \\ 
Transit duration (1.5 to 3.5)~[hr]                                   & $2.557$          &  $\pm 0.004$   &   e  \\ 
Transit ingress or egress duration~[hr]                              & $0.346$          &  $\pm 0.004$  &   e   \\ 
Linear limb-darkening coefficient, $u_1$                             & $0.31$           &  $\pm 0.04$    &  e   \\
Quadratic limb-darkening coefficient, $u_2$                          & $0.354$           &  $-0.05$, $+0.07$  &  e  \\
Orbital inclination, $i$~[deg]                                      & $87.806$           &  $-0.019$, $+0.018$  &  e   \\
Orbital eccentricity, $e$                                         &   $<$0.45~(3$\sigma$) &             &  f \\  
Orbital semimajor axis~[AU]                                        & $0.080$         &  $\pm 0.002$  &   b,e \\
 & & \\
Planet radius, $R_p$~[R$_\oplus$]                                   &  $6.11$         &  $\pm 0.20$            &  b,e \\ 
Planet mass, $M_p$~[$M_\oplus$]                                     & $<$120~(3$\sigma$)   &               &  b,g \\
Planet mean density, $\rho_p$~[g~cm$^{-3}$]                         & $<$3.0~(3$\sigma$)  &     &  b,e,g \\
 & & \\
Sky-projected stellar obliquity, $\lambda$ [deg]                      & $-110$ & $-14$, $+22$    &  h \\
Sky-projected stellar rotation velocity, $v \sin i_\star$~[km~s$^{-1}$] & $5.6$ & $\pm 0.8$       &  h \\ 
Inclination of stellar rotation axis~[deg]                             & $138$ & $\pm 7$         &  i \\
Stellar obliquity, $\psi$ [deg]                                     & $145$ & $-14$, $+9$         &   j  
\enddata

\tablecomments{Each quoted result represents the median of the {\it a posteriori} probability distribution derived from the MCMC, marginalized over all other parameters. The confidence limits enclose 68.3\% of the probability, and are based on the 15.85\% and 84.15\% levels of the cumulative probability distribution.}

\tablenotetext{a}{Based on the SPC analysis of the spectra and the Y$^2$ models, using the gyrochronology age as a constraint (see section \ref{sec:stellarcharac}).}

\tablenotetext{b}{Based on the SPC analysis of the spectra and the Y$^2$ models (see section \ref{sec:stellarcharac}).  The stellar density is given in units of $\rho_{\odot} = 1.408$~g~cm$^{-3}$.}

\tablenotetext{c}{Based on the periodogram of the {\it Kepler} photometric time series (see section~\ref{sec:gyro}).}

\tablenotetext{d}{Based on the Keck/HIRES spectrum (see section~\ref{sec:gyro}).}

\tablenotetext{e}{Based on the analysis of the transit light curves (see section \ref{sec:trans}).}

\tablenotetext{f}{Based on the combination of transit parameters, orbital period, and mean stellar density (see section~\ref{sec:ecc}).}

\tablenotetext{g}{Based on the analysis of the Keck radial velocities, assuming zero eccentricity (see section \ref{sec:RV}).}

\tablenotetext{h}{Based on the analysis on the RM effect (see section~\ref{sec:RM}).}

\tablenotetext{i}{Based on the combination of $P_{\rm rot}$, $R_\star$, and $v\sin i_\star$ (see section~\ref{sec:incli}).}

\tablenotetext{j}{Based on the analysis of the RM effect and starspot-crossing events (see section~\ref{sec:spotlambda}).
\vspace{0.1in}}

\end{deluxetable*}

\subsection{Radial velocity analysis}
\label{sec:RV}

For the radial-velocity analysis, we used the 7 Keck radial velocities obtained before the transit night.  We also took the mean of all the pre-transit radial velocities from the night of 2011~August~20/21, and treated this mean velocity as a single additional data point. A planet somewhat smaller than Jupiter in a 10-day orbit should induce a radial velocity signal with a semiamplitude of order tens of~m~s$^{-1}$.  For a chromospherically quiet star, eight radial velocities with a precision of a few m~s$^{-1}$ would have been enough to determine the mass of the planet. However, in the case of Kepler-63, we expect spurious radial velocities of order tens of~m~s$^{-1}$, the product of $v\sin i_\star$ and the fractional photometric variability. This greatly complicates the mass determination.

\begin{figure}[ht]
\begin{center}
\leavevmode
\hbox{
\epsfxsize=3.5in
\epsffile{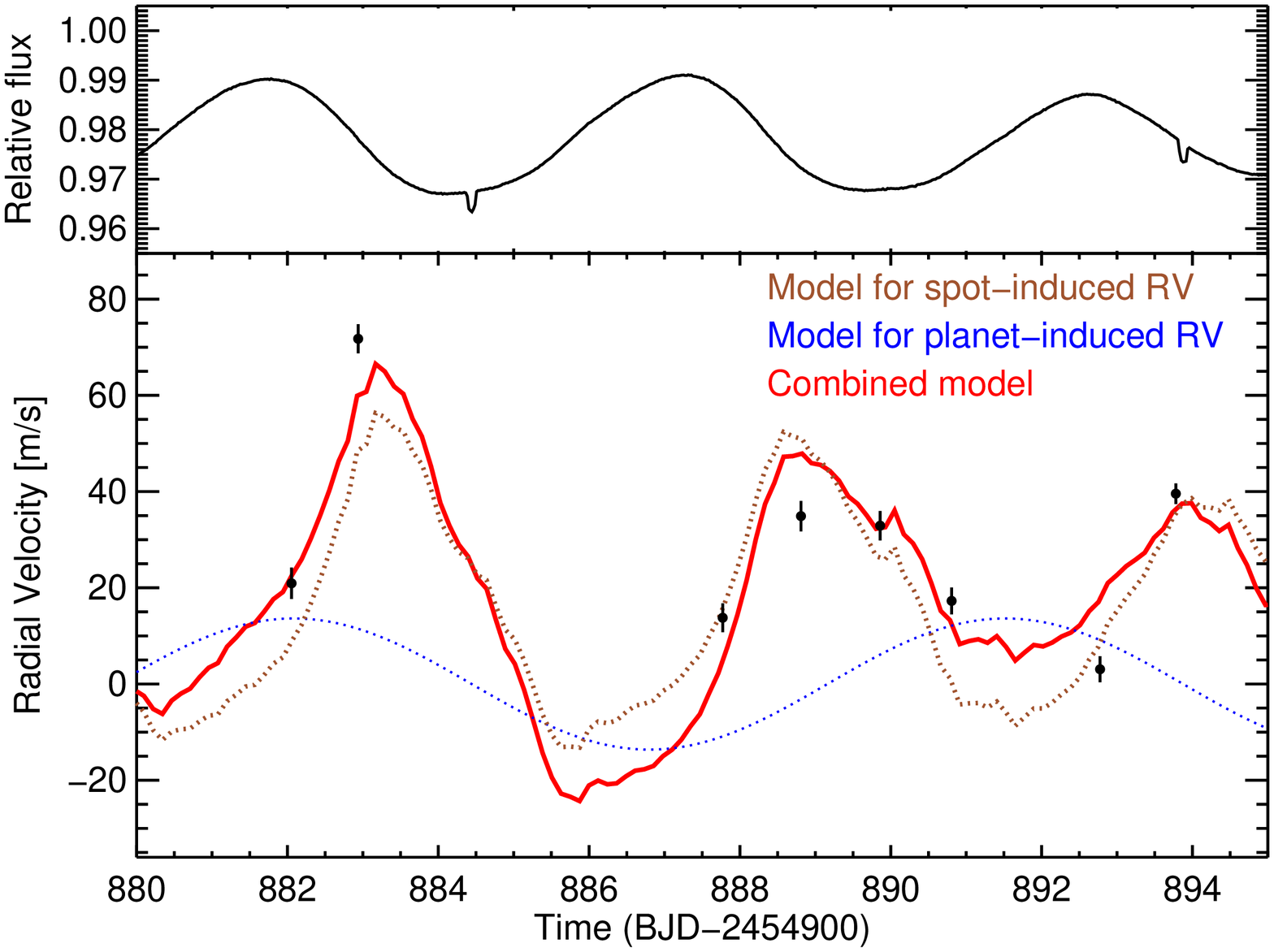}}
\end{center}
\vspace{-0.1in}
\caption{ {\bf Radial-velocity analysis.} {\it Top.}---Relative stellar flux of Kepler-63 during the radial velocity observations. {\it Bottom.}---The black dots represent the radial velocity observations, with vertical bars indicating the internally
estimated measurement uncertainties (with no ``jitter'' term added). The red line represents the optimized model, which is the sum of a sinusoidal function representing the planetary signal (blue line) and the $FF'$ model representing the spurious radial velocity due to rotating starspots (brown line).}
\label{fig:RV}
\vspace{0.1in}
\end{figure}

The top panel of Figure \ref{fig:RV} shows that the stellar flux variations during the radial velocity observations were approximately sinusoidal. The simplest explanation is that a single large dark spot was always on the visible side of the star. In such configurations, it is possible to estimate the spurious radial velocity $V_{\rm spot}$ due to starspots using the so-called $FF'$ method of Aigrain et al.\ (2012). Those authors showed that the spurious radial velocity can be approximated by a function of the normalized stellar flux $f(t)$ and its derivative,
\begin{equation}
V_{\rm spot} (\epsilon, \kappa \delta V_c) =
\dot{f}(t)\left[1-f(t)\right] \frac{R_\star}{\epsilon} + \left[1-f(t)\right]^2 \frac{ \kappa \delta V_c}{\epsilon},
\end{equation}
where the normalization is such that the $f(t) = 1$ level is one standard deviation above the maximum observed flux. There are two free parameters: $\epsilon$ is the relative loss of light due to the spot if it were situated at the center of the stellar disk; and $\kappa \delta V_c$ specifies the alteration of the convective blueshift due the spot. With this ingredient, our model for the radial velocity signal was
\begin{equation}
  V_{\rm calc} = -K \sin{\left[ n(t-t_c) \right]} -
            V_{\rm spot} (\epsilon, \kappa \delta V_c) + \gamma.
\end{equation}
In this formula we have assumed a circular orbit for the planet, with $K$ being the planet-induced radial velocity semiamplitude, and $t_c$ the time of transit. The parameter $\gamma$ represents a constant offset.
The mean motion $n$ is defined as $2\pi/P_{\rm orb}$. Both $P_{\rm orb}$ and $t_c$ are known precisely
from the transit analysis. 

We optimized the model parameters through a standard least-squares fit. Figure \ref{fig:RV} displays the data and the optimized model. Given the typical measurement uncertainty of $2.5$~m~s$^{-1}$, the minimum $\chi^2$ was 78.6 with 4 degrees of freedom. This poor fit is at least partly due to the simplicity of the spot model (which assumes that there is only one small spot on the surface).  By adding a ``jitter'' term of $12.5$ m~s$^{-1}$ in quadrature to the measurement uncertainties we obtain $\chi^2=N_{\rm dof}$.  We use this jitter term and an MCMC algorithm to determine credible intervals for the model parameters, from which we derive a planet mass of $M_p = 45 \pm 26~M_\oplus$.  The 3$\sigma$ upper bound is 120~$M_\oplus$.

Given the severe limitations of this analysis---the weak detection, the imperfect fit of the $FF'$ model, and the assumption of zero eccentricity---we do not claim to have detected the radial-velocity signal due to the planet.  Rather, we interpret the results as a coarse upper bound on the mass of the transiting object, placing it within the planetary regime.

\section{Sky-projected obliquity from the RM effect}
\label{sec:RM}

\begin{figure*}[ht]
\begin{center}
\leavevmode
\hbox{
\epsfxsize=5.8in
\epsffile{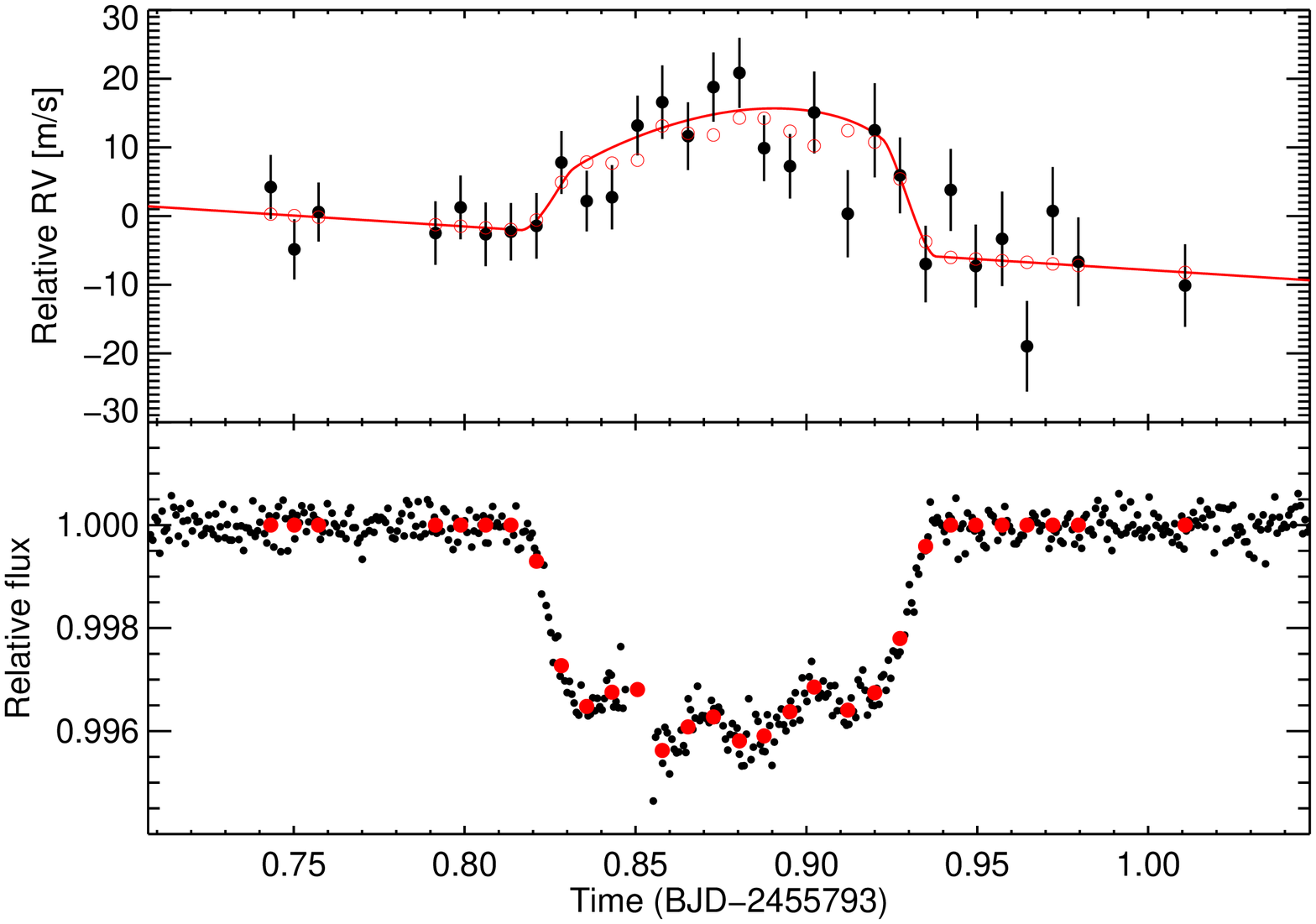}}
\end{center}
\vspace{-0.1in}
\caption{{\bf Evidence for a high obliquity based on the RM effect.}
{\it Top.}---The solid dots are the measured radial velocities. 
The signal is a redshift throughout the transit, 
applying a high obliquity, as opposed to
the red-then-blue signal of a low-obliquity system.
The open dots represent the
best-fitting model. The red curve shows a model with
the same geometric parameters but with the
loss of light appropriate for a spotless star,
to illustrate what one might have observed in the absence of starspots.
{\it Bottom.}---Transit observations in SC mode obtained with \textit{Kepler}. Black dots represent the data and red dots represent the binned light curve used to model the RM effect.}
\label{fig:RM}
\vspace{0.1in}
\end{figure*}

The RM effect is more easily detected than the orbital motion, mainly because the timescale of the RM effect is much shorter than the rotation period of the star, allowing a clean separation between the RM effect and the spurious starspot-induced radial velocities (see, e.g., Gaudi \& Winn 2007). Prior to the Keck observations we had performed enough spot modeling to be able to predict that the anomalous Doppler shift would be a pure redshift throughout the transit. The results of the RM observations, displayed in Figure~\ref{fig:RM}, confirmed this prediction. However, the {\it Kepler} photometry of the same transit (shown in the bottom panel of Figure~\ref{fig:RM}) revealed at least three spot-crossing events, complicating the modeling.

To model the RM effect, one needs the usual parameters describing the loss of light as well as 4 additional parameters: $v\sin i_\star$ and $\lambda$ to describe the amplitude and shape of the signal, and a slope $\dot{\gamma}$ and offset $\gamma$ to account for the orbital motion of the star. Usually the loss of light is computed based on the transit parameters $b$, $R_{\rm s}/a$, $R_{\rm p}/R_\star$ and the time of transit.  Here, given the presence of spot-crossing anomalies, we chose to take the loss of light directly from the {\it Kepler} photometric time series, since this naturally takes the anomalies into account, and the cadence and precision are more than sufficient for our purpose. To obtain the loss of light corresponding to each point in the RV time series, we averaged the corresponding {\it Kepler} photometric data points. The final averaged light curve is shown in the lower panel of Figure~\ref{fig:RM} as the sequence of red dots.

The anomalous RV was then computed with the formulas of Hirano et al.\ (2011), using the planet position and loss of light as inputs.  This code takes into account the effects of macroturbulence, pressure broadening, and instrumental broadening.  Model fitting and parameter estimation were performed using the protocols of Albrecht et al.\ (2012). In particular we imposed Gaussian priors on $T_{14}$ and $T_{12}$, based on the parameters reported in Table \ref{tbl:params}.  We also used the parameter combinations $\sqrt{v\sin i_\star}\sin \lambda$ and $\sqrt{v\sin i_\star}\cos \lambda$ rather than $v\sin i_\star$ and $\lambda$, to minimize correlations. The uncertainty in each RV data point was taken to be the quadrature sum of the internally estimated uncertainty and 4.8 m~s$^{-1}$, the value for which $\chi^2=N_{\rm dof}$. The result for the sky-projected obliquity is $\lambda = -110_{-14}^{+22}$~deg.

There are some other interesting results of this analysis. We find the projected rotation speed to be $v \sin i_\star = 5.6 \pm 0.8 \, \kms$, in agreement with the value obtained from the basic spectroscopic analysis (see section~\ref{sec:stellardim}).  The result for the out-of-transit velocity slope $\dot{\gamma} = -30 \pm 15$ m~s$^{-1}$~day$^{-1}$ can be translated into an estimate of the velocity semiamplitude due to the planet, using the orbital period and assuming a circular orbit. The result is $K_{\rm RM} = -\dot{\gamma} P_{\rm orb}/2\pi = 40 \pm 20$ m~s$^{-1}$. This slope is compatible within the uncertainties with the measured $K$ from the RV analysis (approximately 15 m~s$^{-1}$), although the uncertainties are large, and the effects of spots were not taken into account in this determination of $K_{\rm RM}$. In this case, by chance, the transit happened about a quarter of a rotation cycle before a flux minimum, which is when the spot-induced spurious acceleration is expected to be small.

\section{Obliquity measurement from starspots}
\label{sec:obliquity}

\subsection{Stellar inclination from $v\sin i_\star$}
\label{sec:incli}

We combined the values of the rotation period $P_{\rm rot}$, stellar radius $R_\star$, and sky-projected stellar rotation velocity $v\sin i_\star$, to obtain $\sin i$, the inclination of the stellar rotation axis with respect to the line of sight. Based on the values given in Table~\ref{tbl:params}, the stellar rotation velocity is $v = 2 \pi R_\star/P_{\rm rot} = 8.4 \pm 0.2$~km~s$^{-1}$. This is significantly larger than $v\sin i_\star = 5.6 \pm 0.8$~km~s$^{-1}$ obtained from the analysis of the RM effect, implying $\sin i_\star <1$.  The implied stellar inclination angle is either $42\pm 7$~deg or $138\pm 7$~deg. As we will see in the next section, the latter value of the stellar inclination is favored. Since Kepler-63b is transiting with an orbital inclination of $87.81\pm 0.02$~deg, this simple analysis demonstrates the star has a high obliquity, independently of the RM effect.

We used the marginalized posterior for $\lambda$ obtained in the last section, as well as those for the stellar and orbital inclinations (Table \ref{tbl:params}), to obtain the true obliquity $\psi$. Using the formula from Fabrycky \& Winn (2009), the result is $\psi = 104_{-14}^{+9}$~deg.

\subsection{Sky-projected obliquity from spot-crossing anomalies}
\label{sec:spotlambda}

In principle the obliquity of the system is also encoded in the pattern of photometric variability and spot-crossing anomalies, but in this case the anomalies are so numerous that the pattern has proven difficult to interpret unambiguously. Rather than attempt a rigorous independent determination of the obliquity, we discuss here a starspot model that at least demonstrates the compatibility between the photometric variability, the starspot-crossing events, and the preceding results for $\lambda$ and $i_\star$.

\begin{figure*}[ht]
\begin{center}
\leavevmode
\hbox{
\epsfxsize=7.0in
\epsffile{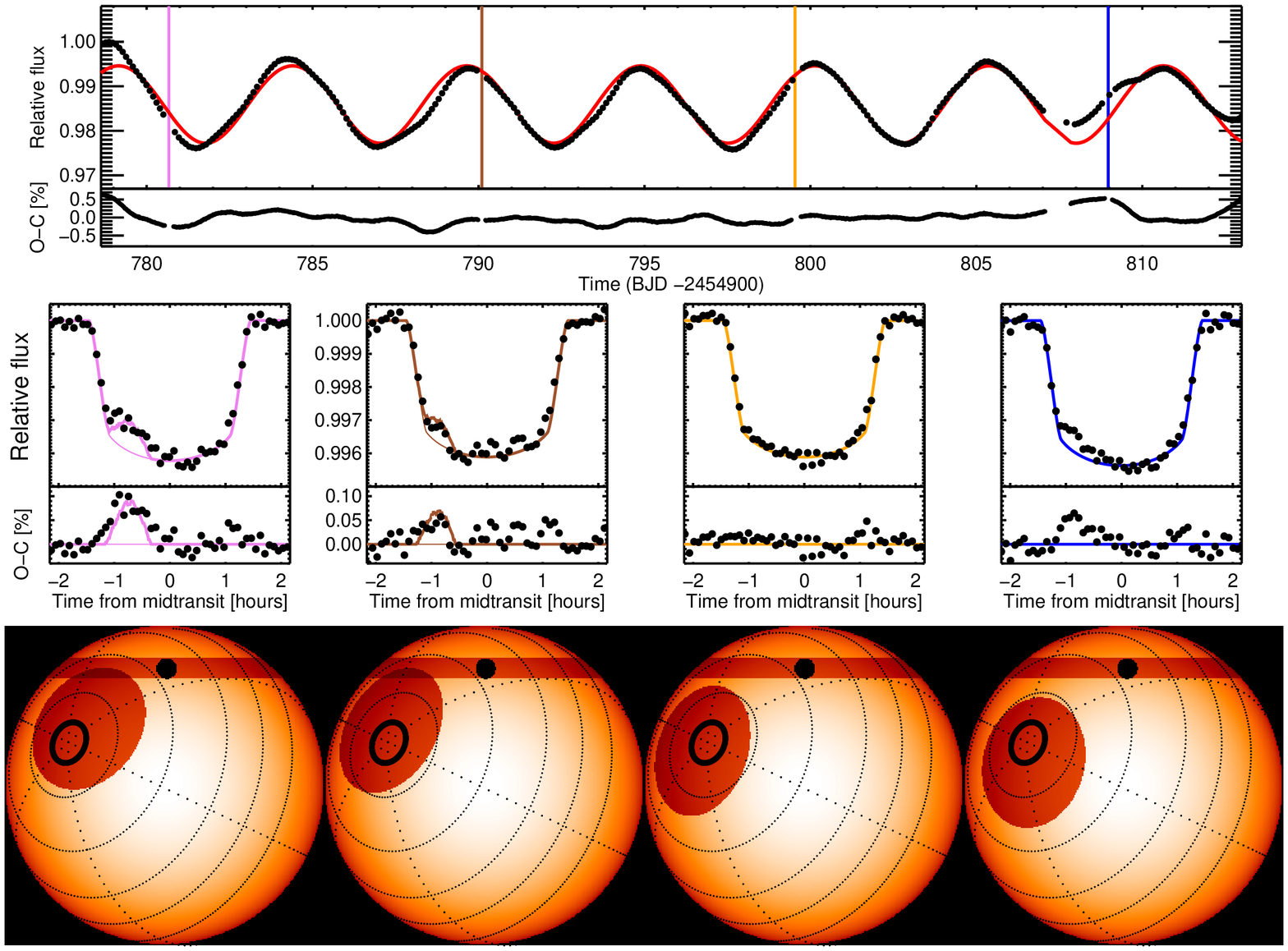}}
\end{center}
\vspace{-0.1in}
\caption{ {\bf Evidence for a large obliquity from a single-spot model.} {\it Top.}---Relative flux of Kepler-63 (black dots) over a time range spanning
four transits. The transit times are marked with vertical lines.
The red line represents the best-fitting model with a single starspot.
{\it Center.}---Transit light curves, with 5~min sampling.
The thick curves represent the best-fitting model with a single starspot;
the thin curves show the model with the spot darkening set equal to zero.
The model accounts for the two largest spot-crossing anomalies but the residuals indicate that more spots are present.
{\it Bottom.}---Locations of the spot, transit chord, and planet at midtransit,
according to the best-fitting model.
}
\label{fig:fullspot}
\vspace{0.2in}
\end{figure*}

We focused attention on a time interval when the overall photometric variability seemed relatively simple: a nearly sinusoidal pattern with peak-to-peak amplitude of about 1.5\%. This interval spans four consecutive transits, specifically epochs 71-74 (see Figure~\ref{fig:fullspot}). A large and long-duration spot-crossing anomaly is seen in the first half of the first transit. We proceeded by assuming this is the same spot that is producing the quasi-periodic stellar flux variation, and attempted to model all of the data under this premise.

The orientation of the star was parameterized by the sky-projected obliquity $\lambda$ and the inclination angle $i_\star$. The rotation period was a free parameter, which was tightly constrained by the quasi-periodic variability. A Gaussian prior constraint was imposed on $v \sin i_\star$ based on the results of section~\ref{sec:incli}.

We modeled the out-of-transit variability using the Dorren (1987) equations for the loss of light due to a starspot. We fixed the limb-darkening coefficient to a value of 0.56, which provides the best fit to the light curve constructed in section~\ref{sec:trans}. The spot's brightness contrast relative to the photosphere was taken to be a constant over the interval of the observations. The observed phase of the out-of-transit variability specifies the spot longitude, and the stellar rotation period is also well constrained. Therefore, given particular choices for the orientation of the star and the spot latitude, we could calculate the location of the spot at any time, including the times of the four transits.

The transits were modeled using the geometric transit parameters from section~\ref{sec:trans}, and a pixelated stellar disk. At any particular time we computed the sum of the intensities of all the pixels, some of which were darkened by the spot or hidden by the planet (Sanchis-Ojeda et al.\ 2011).

The best-fitting value for $\lambda$ was $-115^{\circ}$, in agreement with the result based on the RM effect. The model also prefers $i_\star = 135^{\circ}$, selecting one of the two values for the inclination that were allowed by the analysis in section~\ref{sec:incli}. This constraint arises from the requirement that the spot must cross the transit chord before its closest approach to the center of the stellar disk. In the optimized model, the spot is large and resides near one of the rotation poles.

The simple one-spot model is therefore compatible with the overall photometric variability and the largest spot anomaly. The smaller spot anomaly during the second transit can also be attributed to this spot, as illustrated in Figure~\ref{fig:fullspot}. Certainly, though, this model does not capture all of the sources of photometric variability: there are at least six other smaller anomalies that are not well-fitted, and which it does not seem worthwhile to try and model. The large anomaly in the fourth transit agrees in phase with the two other explained anomalies. It is possible that the same spot is responsible for this anomaly, if the spot has an irregular shape.

\section{Starspot characteristics and magnetic cycles}
\label{sec:magnetic}

\begin{figure*}[ht]
\begin{center}
\leavevmode
\hbox{
\epsfxsize=4.6in
\epsffile{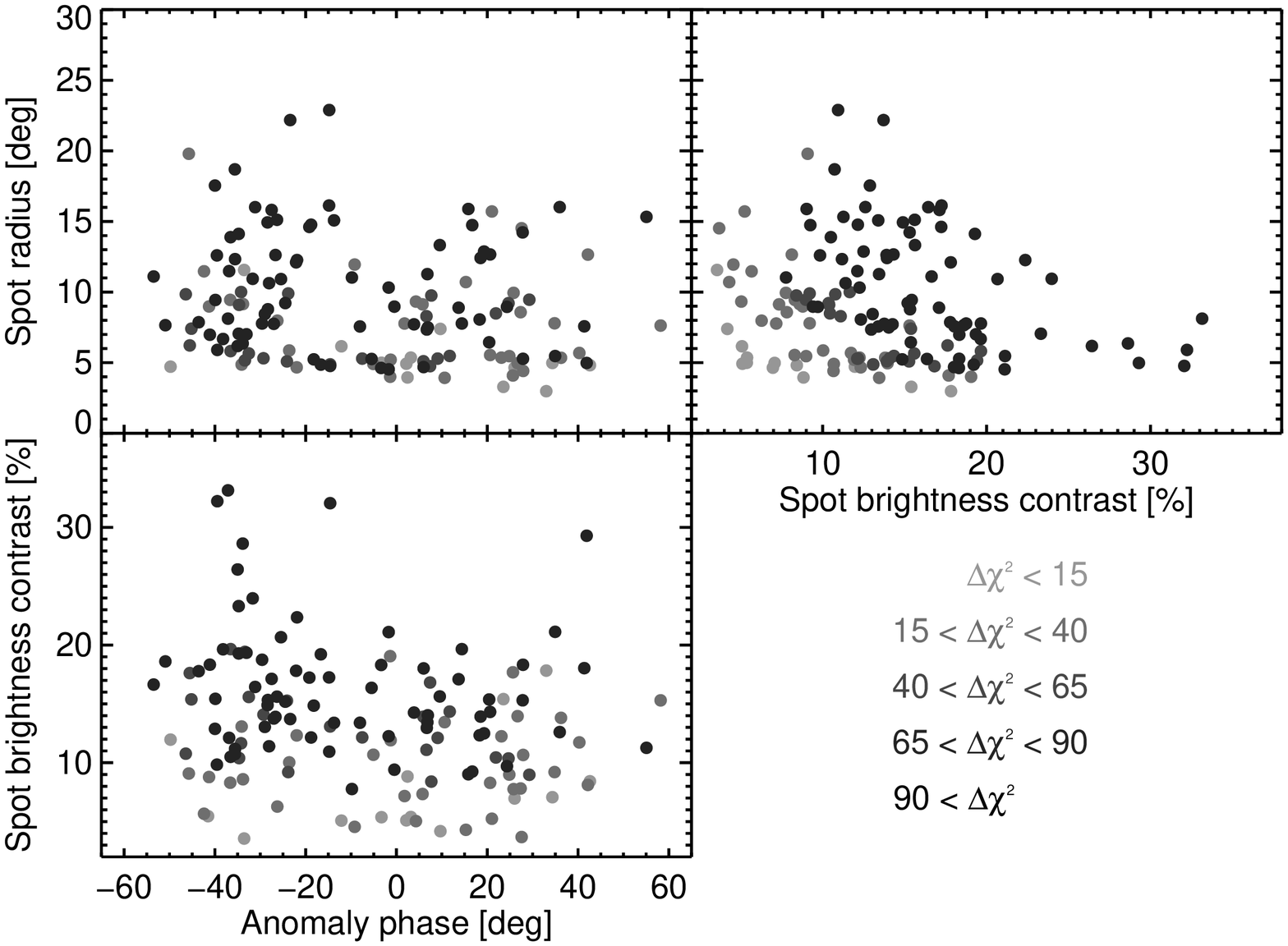}}
\end{center}
\vspace{-0.1in}
\caption{ {\bf Characteristics of the spot-crossing anomalies in Kepler-63b. } The figure shows the best-fitting values of the angular radius,
brightness contrast, and phase of the 145 spot-crossing anomalies that were
identified.
Darker dots represent more significant detections. }
\label{fig:anom}
\vspace{0.1in}
\end{figure*}

\begin{figure*}[ht]
\begin{center}
\leavevmode
\hbox{
\epsfxsize=5.2in
\epsffile{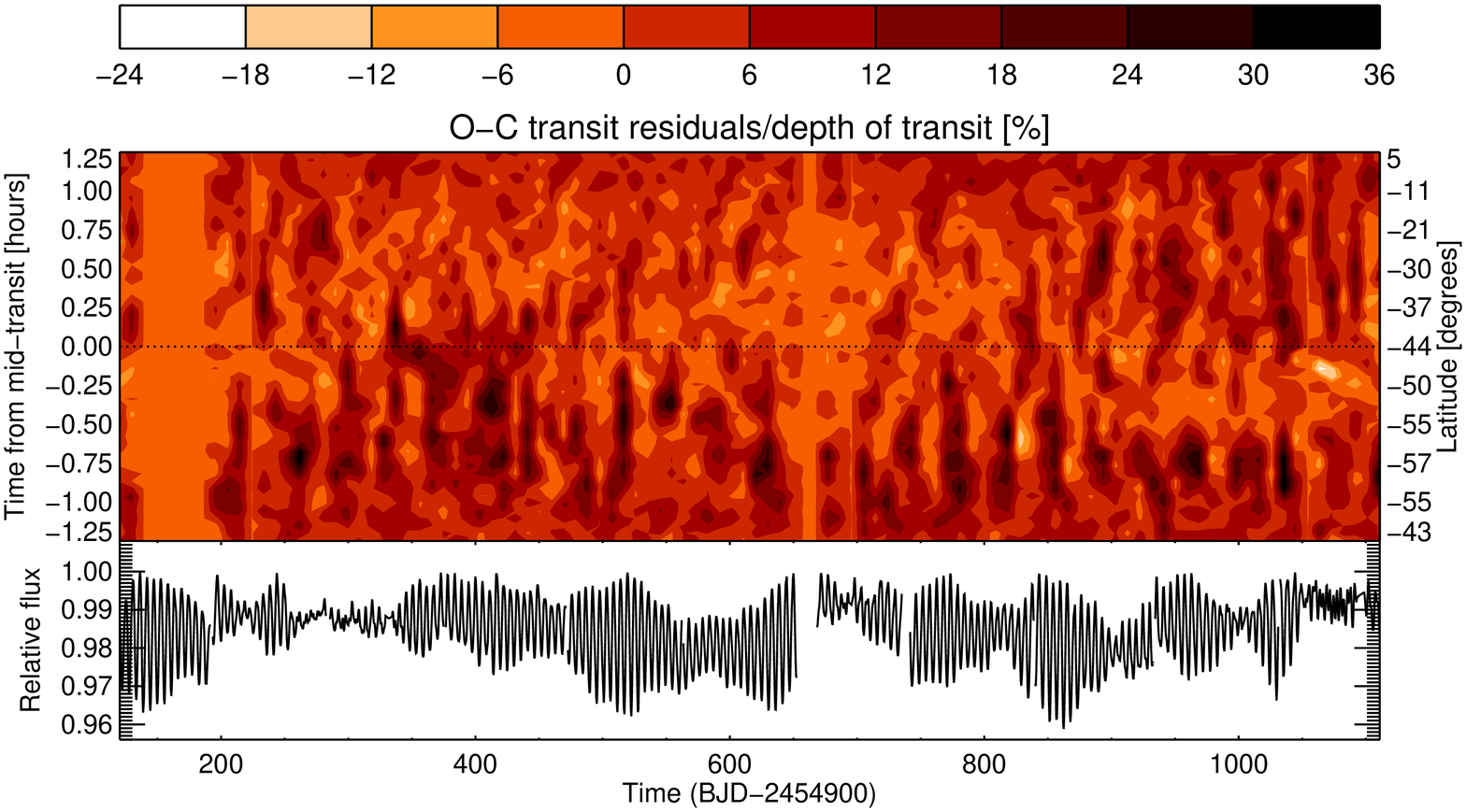}}
\end{center}
\vspace{-0.1in}
\caption{ {\bf Spot latitude evolution.} {\it Top.}---Residuals between the
SC transit data and the best-fitting model, after dividing
by the transit depth. The dark regions represent spot-crossing anomalies.
The vertical axes indicate the time relative to the midtransit time (left)
and the corresponding latitude of the position of the planet (right).
The high-latitude spot is crossed every fourth transit.
Some activity is also seen at lower latitudes.
{\it Bottom.}---Relative flux variation over the same time interval.
For this plot the data from each {\it Kepler} quarter were normalized
by the maximum quarterly flux.}
\label{fig:butter}
\vspace{0.1in}
\end{figure*}

As discussed in section~\ref{sec:trans}, we visually identified 145 spot-crossing anomalies. To study the position, sizes, and temperatures of the spots, we modeled the individual spot-crossing anomalies with the same pixelated spot model discussed in section~\ref{sec:spotlambda}.  The parameters describing each spot-crossing event were the spot's angular radius and brightness contrast, as well as the timing of the event, which we express as an ``anomaly phase'' ranging from $-90^\circ$ (ingress) to $90^\circ$ (egress). The other transit parameters, including the spot coverage factor $\epsilon$, were taken from the analysis of section~\ref{sec:trans}. To evaluate the significance of detection of the anomalies, we used $\Delta\chi^2$ between the best-fitting spot model and the best-fitting spot-free model.

There is a degeneracy between the modeled position and radius of a spot, because we lack the precision to measure the impact parameter between the planet and spot (and there is anyways no reason to think the spot is perfectly circular). To avoid this degeneracy, we assumed that the planet passes through the center of the spot.  It should be understood, then, that the ``spot radius'' in our model is really a measure of the length of the intersection between the spot and the transit chord.

Figure~\ref{fig:anom} shows the results. Unsurprisingly, the significance of detection increases with the size and the brightness contrast of the spots. The anomalies that appear in the first half of the transit (negative anomaly phase) are generally more significant and more abundant.  The typical spot radius is $10^\circ$, and the largest spots have a radius of 15--20$^\circ$.  The typical brightness contrast is 15-20\%. To produce a 20\% brightness contrast in the {\it Kepler} bandpass (450--850~nm) would require an effective temperature approximately 300~K lower than the photosphere, assuming blackbody spectra.

Figure \ref{fig:butter} shows the time evolution of the spot-crossing events. To create this figure we subtracted a model for the loss of light due only to the planet, and then divided the residuals by the transit depth. The normalized residuals were then plotted as a function of both time (horizontal axis) and phase within the transit (vertical axis). Given the known orientation of the star, we can also translate the phase within the transit into a stellar latitude (second vertical axis). The transit chord spans latitudes from $-60^\circ$ to $5^\circ$, that is, a large portion of the southern hemisphere. Note also that the relation between the transit phase and the stellar latitude is nonlinear; indeed some latitudes cross the transit chord in more than one location.

Spot-crossing events are visible as dark regions in this plot. Most of the activity is seen in the early portions of the transits. This indicates a long-lived polar active region. After the first few years of observations (starting at around day 875) spot-crossing anomalies began to appear in the second half of the transit, corresponding to lower stellar latitudes.  Anomalies at mid-transit were comparatively rare, especially once the second half of the transit chord became active.  Perhaps this is a sign that active regions tend to be segregated in latitude, with some activity at high polar latitudes and some at more equatorial latitudes.

These initial explorations of the spatial distribution of activity on Kepler-63 could be continued in the future by developing a multi-spot model, fitted to both the stellar flux variations and the spot-crossing anomalies (see, e.g., Bonomo \& Lanza 2012, Oshagh et al.\ 2013a). Here we have focused mainly on the anomalies, which provide snapshots of the transit chord every 9.4 days; there are undoubtedly some spots that are missed with this approach (Llama et al.\ 2012).
 
\section{Discussion}
\label{sec:discussion}

\begin{figure*}[ht]
\begin{center}
\leavevmode
\hbox{
\epsfxsize=5.2in
\epsffile{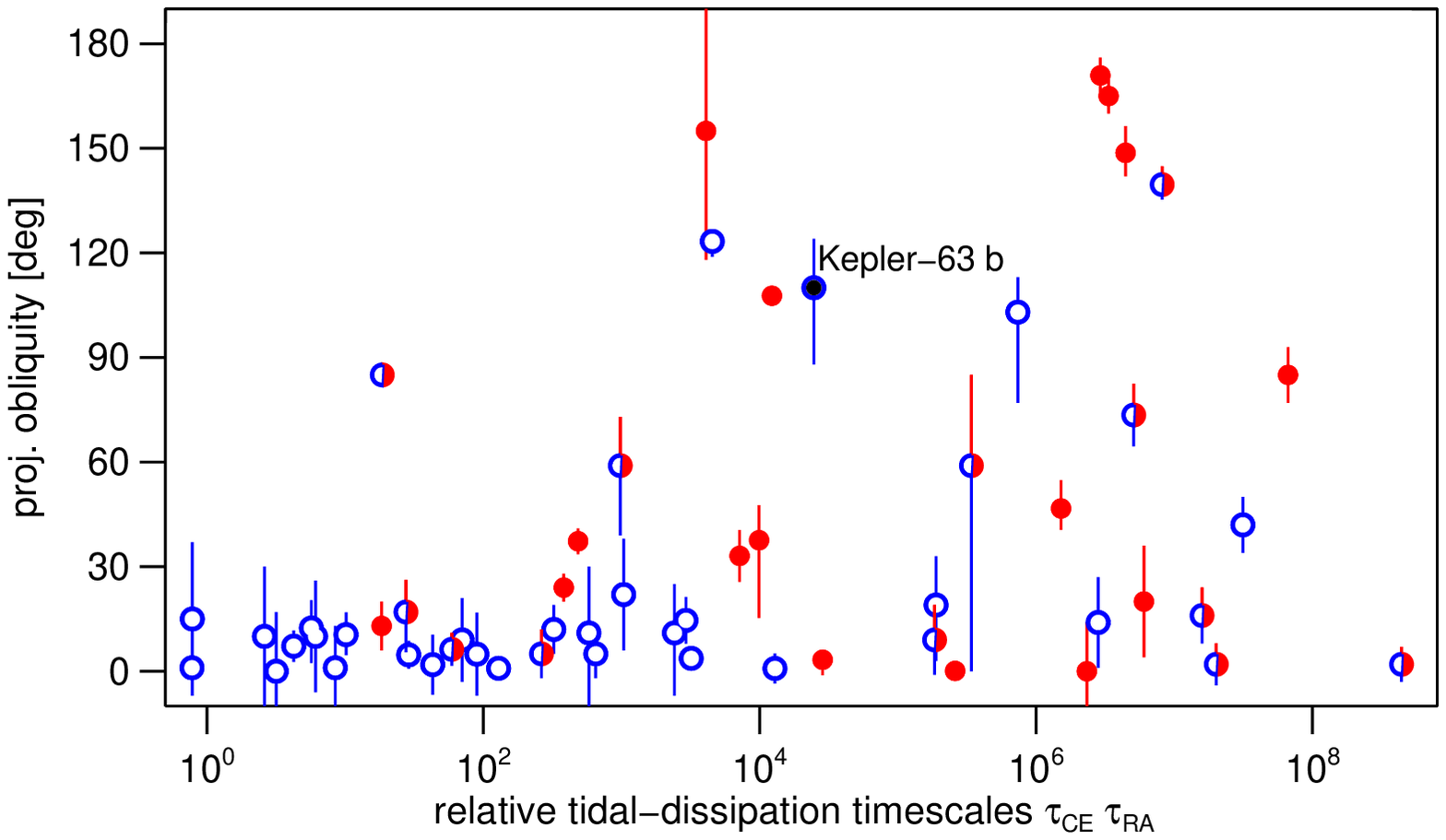}}
\end{center}
\vspace{-0.1in}
\caption{ {\bf Sky-projected obliquities as a function of
relative timescale for tidal dissipation.} See Albrecht et al.~(2012)
for the original figure on which this is based, and for details
on its construction.
In short, the relative tidal dissipation timescale
is assumed to be proportional to $q^2 (a/R_\star)^6$ for stars with
$T_{\rm eff}<6250$~K (blue dots)
and proportional to $q^2(1+q)^{5/6}(a/R_\star)^{8.5}$ for hotter stars (red dots), where $q$ is the planet-to-star mass ratio.
The hot and cool stars are placed on the same scale
using an empirical calibration based on observations of binary-star circularization periods. Dots with both colors
represent stars for which the measured $T_{\rm eff}$ straddles the boundary.
Lower obliquities are seen
in systems with relatively rapid tidal dissipation,
suggesting that tides are responsible for damping stellar obliquities.
Even though Kepler-63 is a cool star which is relatively dissipative,
the orbital distance is large enough that tides are relatively weak,
and the high stellar obliquity fits well with this observed trend.
In addition to Kepler-63 this plot features
new and updated values of $\lambda$ for CoRoT-11b (Gandolfi et al.\ 2012), WASP-19b (Tregloan-Reed et al.\ 2013), WASP-32b and WASP-38b (Brown et al.\ 2012) and
HAT-P-17b (Fulton et al.\ 2013).  }
\label{fig:simon}
\vspace{0.1in}
\end{figure*}

In this paper we have presented Kepler-63b, a giant planet transiting a star on an orbit that is highly inclined with respect to the stellar equator. On the one hand the star's high levels of chromospheric activity interfered with our ability to characterize the system through transit light curve analysis and radial-velocity monitoring. On the other hand the {\it Kepler} data allowed us to partly correct for the effects of activity; and also to take advantage of the activity to determine the stellar rotation period, explore the spatial distribution of starspots, and perform a consistency check on the stellar obliquity that was determined via the Rossiter-McLaughlin effect.

The measurement of the planet's mass through radial-velocity measurements was unsuccessful because the spurious radial velocities caused by starspots were larger than the planet-induced signal. To measure the mass, a large body of additional radial velocities will be required, in a campaign that is carefully designed to try and separate the effects of rotating starspots and orbital motion.  The information about the general spot characteristics presented in this paper may help in designing such a campaign.

The star's high obliquity corroborates the scenario proposed by Winn et al.\ (2010a) and Albrecht et al.\ (2012) in which hot Jupiters have orbital inclinations that are initially nearly random with respect to the stellar equator, and are eventually damped to low inclinations if the tidal interactions between the star and planet are sufficiently strong.  In this scenario a high obliquity is expected for Kepler-63, because even though the star is relatively cool and has a thick convective envelope (a factor leading to relatively rapid tidal dissipation), the orbital distance is relatively large. To be quantitative we used the metric developed by Albrecht et al.\ (2012), in which binary-star data are used to calibrate tidal dissipation timescales. Figure~\ref{fig:simon} shows that the expected timescale for tidal dissipation for this system is in the regime where random alignment is observed among the other close-in giant planets. The fact that the star is young also helps to understand why it has not yet been realigned (Triaud~2011). This measurement is interesting because among planet-hosting stars with measured obliquities, only HAT-P-11b is comparable to Kepler-63b in size and orbital period, being smaller (4.7~$R_\oplus$) than Kepler-63b but also having a shorter orbital period (4.9~days).

A proposed interpretation for these findings is that hot Jupiters begin far away from the star, beyond the snow line, where it is easier to understand their formation. The initial obliquity of the system is low, as a consequence of the formation of the entire system from a single disk of gas and dust. Then, dynamical interactions such as planet-planet scattering (Rasio \& Ford 1996) or Kozai cycles induced by the influence of a distant companion (Fabrycky \& Tremaine 2007), move the planet into a highly eccentric orbit with a more random orientation.  In this highly eccentric orbit, the planet passes very close to the star, where tidal interactions are significant. This tidal interactions will circularize and shrink the orbit and, if they are strong enough, will realign the spin axis of the star with the orbital angular momentum. In the context of this theory, pursuing obliquity measurements for systems with smaller planets and longer orbital periods is interesting because at a certain point those planets might have been able to form {\it in situ}, leading to an expectation of a population of well-aligned systems.

Rogers et al.\ (2012) have proposed that at least some of the high obliquities might have nothing to do with planet formation {\it per se} but are instead the consequence of reorientation of stellar photospheres due to the redistribution of angular momentum by internal gravity waves. Their theory is applicable to stars with radiative envelopes, and is therefore not applicable to Kepler-63, nor to the other three cool stars with high obliquities that are seen in Figure~\ref{fig:simon}.

In addition to measuring the obliquity of Kepler-63 we have confirmed that the planet is passing in front of a large, dark, persistent spot (or group of spots) located near one of the star's rotation poles.  Such spots are not seen on the present-day Sun, where the spot latitudes follow an 11-year cycle in which they start appearing at medium latitudes (30--40$^\circ$) and end up appearing near the equator (for a review, see Solanki 2003). However, there was previous evidence that polar spots are common around young Solar analogs. This was based on simulations of magnetic activity (Brown et al.\ 2010; Schrijver \& Tittle 2011) as well as empirical evidence from Doppler imaging of young and rapidly rotating stars such as EK Dra (Strassmeier \& Rice 1998). Even though such polar spots were detected in different occasions and with different techniques (Strassmeier et al.\ 1991), and multiple tests were performed to validate the technique (Unruh \& Collier Cameron 1995; Bruls et al.\ 1998), an independent confirmation using a different method was previously lacking. Our study provides further evidence for these types of spots, through a direct method based on periodic occultations of the spots by a planet with a well-understood geometry. The current information gathered about stellar spots on Kepler-63, and future studies that could analyze the information from stellar flux variations, may provide useful information about the activity of young Sun-like stars. It would also be interesting to find additional active stars with transiting planets in the {\it Kepler} database, as a probe not only of stellar obliquities but also starspot characteristics and evolution.

\acknowledgements We thank the anonymous referee for numerous insightful suggestions that led to major improvements in this paper. We also thank Andrew Collier Cameron, Bryce Croll, and Benjamin Brown for helpful discussions, and the entire {\it Kepler} team for the success of the mission. R.S.O.\ and J.N.W.\ acknowledge NASA support through the {\it Kepler} Participating Scientist program. {\it Kepler} was competitively selected as the tenth Discovery mission. Funding for this mission was provided by NASA's Science Mission Directorate. The data presented in this article were obtained from the Mikulski Archive for Space Telescopes (MAST). STScI is operated by the Association of Universities for Research in Astronomy, Inc., under NASA contract NAS5-26555. Support for MAST for non-HST data is provided by the NASA Office of Space Science via grant NNX09AF08G and by other grants and contracts. J.A.C.\ acknowledges support by NASA through a Hubble Fellowship (grant HF-51267.01-A). RID is supported by the NSF-GRFP (DGE-1144152). J.A.J.\ is supported by generous grants from the Alfred P.\ Sloan Foundation and the David and Lucile Packard Foundation. T.L.C., W.J.C., and G.R.D.\ acknowledge the support of the UK Science and Technology Facilities Council (STFC). Funding for the Stellar Astrophysics Centre is provided by The Danish National Research Foundation (Grant agreement DNRF106). This research was partly supported by the ASTERISK project (ASTERoseismic Investigations with SONG and {\it Kepler}) funded by the European Research Council (Grant agreement no.\ 267864). G.T.\ acknowledges partial support for this work from NSF grant AST-1007992.

\end{document}